\documentclass[smallextended]{svjour3}       

\smartqed

\usepackage{graphicx}
\usepackage{csquotes}
\usepackage[table]{xcolor}
\usepackage[square,numbers]{natbib}
\usepackage{amsmath}
\usepackage{amssymb}
\usepackage{mathptmx}
\usepackage{epigraph}
\usepackage{physics}
\usepackage[utf8]{inputenc}

\begin{document}

\title{Space-time-hap: a coordinate system for the multiverse and its application to show that free choice is observer-dependent}
\titlerunning{Space-time-hap}


\author{Ghislain Fourny}


\institute{G. Fourny \at
              ETH Z\"urich, Switzerland \\
              Department of Computer Science \\
              Stampfenbachstrasse 114 \\
              8092 Z\"urich, Switzerland \\
              \email{ghislain.fourny@inf.ethz.ch}\\
              ORCID 0000-0001-8740-8866\\
}

\date{May 9, 2022}

\maketitle

\begin{abstract}

This note proposes a paradigm and coordinate system that extends flat, four-dimensional Minkowski spacetime to a broader framework that identifies an event not only in space and in time, but also in terms of possible world, with a third category of coordinates called ``hap'' modelling contingency and counterfactuals. Semantically, hap is based on a Bohmian-like configuration space, in which initial conditions at a specific moment in time uniquely identify a trajectory. This framework can be used to support reasonings that rigorously distinguish between causal dependencies (through time), statistical dependencies (through space) and counterfactual dependencies (through hap). As an example, we use this framework to show that the assumption of free choice is not absolute, but rather depends on the chosen frame of reference: while Alice may see a choice made freely, which is formally a unilateral, single-coordinate translation in hap, Bob sitting in another reference frame might see this same choice not made freely and observe instead a translation in hap jointly across multiple coordinates, which indicates a counterfactual dependency. The framework is agnostic regarding which flavour of the De-Broglie-Bohm theory is considered, as it only makes general assumptions on the set of trajectories that are commonly accepted in the Bohmian mechanics community. It is also general enough to support reasoning on theories for which the separation between possible worlds might also depend on the frame of reference, i.e., two observers might disagree on whether two spacetimehap events happen in the same world or not.



\end{abstract}

\bigskip

\section{Introduction}
In this research note, I would like to share and explain an idea that has been on my mind for a few years now. I believe it is possible and useful to extend spacetime \cite{Einstein1905} to a wider framework that I suggest to call ``spacetimehap''.

The goal of the hap part is to formally capture counterfactuals \cite{Lewis1973}\cite{Stalnaker1968}\cite{Stalnaker1972}, which are, from the perspective of a given observer located in a specific possible world \cite{Kripke1963} \cite{Kripke1965}, events that do not happen, but could have happened. Counterfactuals are not part of relativity theory, but they are a core part of quantum theory \cite{Fuchs2013} in particular when considering mixed states and ensembles in statistical physics. Counterfactuals are explicitized in the De Broglie-Bohm interpretation of quantum physics via Bohmian trajectories \cite{Bohm1987}\cite{Teufel2009}. In fact, Bohmian trajectories form a starting point in the construction of spacetimehap, although the introduced framework is not restricted to Bohmian mechanics.

\section{Special relativity and causality}

In special relativity \cite{Einstein1905}, an event with a well-defined position in spacetime is identified using a four-coordinate system in $\mathbb{R}^4$. More precisely, an event has three coordinates of space and one coordinate of time. Mathematically, this is known as Minkowski spacetime \cite{Minkowski1908}.

Minkowski spacetime is a special kind of manifold. First, it is pseudo-Euclidian because the time dimension is special: the squared distance between two events ``across time''\footnote{timelike-separated, as we will see shortly} has an opposite sign to that ``across space''\footnote{spacelike-separated, as we will see shortly}. If we take a vector $\textbf{w}$ with coordinates $(t, x, y, z)$ (seen by, say, Alice), then\footnote{For simplicity we shall normalize by taking $c=1$. Also, there another equivalent convention with $x^2+y^2+z^2  - t^2$.}

$$\|\textbf{w}\|^2 = \textbf{w}^\mu \textbf{w}_\mu = t^2 - x^2 - y^2 - z^2$$

where we see it can be positive or negative.

Second, Minkowski spacetime is a flat manifold, meaning that the metric is the same everywhere, which simplifies a lot of things: there is no curvature, and there is no torsion. In the above norm computation we implicitly took the metric to be everywhere the bilinear form defined, in any inertial reference frame, as:

$$g=\left[
\begin{array}{cccc}
1 & 0 & 0 & 0 \\
0 & -1 & 0 & 0 \\
0 & 0 & -1 & 0 \\
0 & 0 & 0 & -1
\end{array}
\right]$$

That is, the Minkowski squared norm is more explicitly:

$$\textbf{w}^\mu \textbf{w}_\mu = \textbf{w}^\mu g_{\mu\nu} \textbf{w}^\nu = t^2 - x^2 - y^2 - z^2$$

If another observer Bob moves at constant speed $\textbf{v}=(v_x, v_y, v_z)$ from Alice, then the corresponding change of coordinates to those seen by this other observer is given by a Lorentz transformation. In this note, we denote it with a Lorentz boost $B_\textbf{v}$:

$$(t', x', y', z')=B_v(t, x, y, z)$$

where (t', x', y', z') are the coordinates seen in Bob's reference frame.

In more details, the Lorentz boost expressed in any reference frame can be represented with the matrix (again, we take $c=1$):

$$\left[
\begin{array}{cccc}
\gamma & -\gamma v_x & -\gamma v_y & -\gamma v_z\\

- \gamma v_x & 1 + (\gamma-1)\frac{v_x^2}{v^2}& (\gamma-1)\frac{v_x v_y}{v^2}&(\gamma-1)\frac{v_x v_z}{v^2}\\

- \gamma v_y & (\gamma-1)\frac{v_y v_x}{v^2}& 1+(\gamma-1)\frac{v_y^2}{v^2}&(\gamma-1)\frac{v_y v_z}{v^2}\\

- \gamma v_z & (\gamma-1)\frac{v_z v_x}{v^2}&(\gamma-1)\frac{v_z v_y}{v^2}& 1+(\gamma-1)\frac{v_z^2}{v^2}
\end{array}
\right]$$

where

$$v^2 = v_x^2 + v_y^2 + v_z^2$$

and

 $$\gamma=(1-v^2)^{-\frac{1}{2}}$$

The squared distance between two events (the square of the norm formed by the vector joining them) is independent from the frame of reference and is thus preserved by a Lorentz transformation, meaning in particular that the \emph{sign} of the metric is preserved. We have

$$ \textbf{w}^\mu \textbf{w}_\mu = t'^2 - x'^2 - y'^2 - z'^2= t^2 - x^2 - y^2 - z^2$$

if the coordinates differ by a Lorentz boost. This property is called Lorentz invariance.

With our convention, when the sign of the squared distance between two events is positive, the events are said to be timelike separated. If the squared distance is negative, then the events are said to be spacelike separated. And timelike or spacelike separation does not depend on the observer.

When two events $E_1$ and $E_2$ are timelike-separated, all observers see them happen in the same order, meaning that if $t_1<t_2$ for some observer, then another observer for whom they happen at times $t'_1$ and $t'_2$ relative to them will see $t'_1<t'_2$ as well and all observers thus agree that $E_1$ precedes $E_2$: $E_1 E_2$ is future-directed timelike separated.

When two events are spacelike-separated, however, observers may disagree on the order of events. While for Alice $t_1<t_2$, Bob might have $t'_2<t'_1$. This was a major change compared to Newtonian physics, because it means that time is not absolute.

The special case when the distance is 0 corresponds to lightlike (or null) separation and to a moving photon.

The notion of timelike-separation vs. spacelike-separation is fundamental in the study of causality \cite{Vilasini2022}. Causal dependency graphs, in the sense of relativistic causality, can directly be derived and abstracted away from the spacetime structure: there is a directed edge from an event $E$ to an event $F$ in the causal dependency graph exactly when they are timelike (or lightlike) separated and $E$ happens before $F$ for all observers. It is common to draw the so-called transitive reduction of a causal dependency graph to keep the display of the causal structure simple.

Note that relativistic causality as used in this note should be contrasted with causality in causal models \cite{Pearl2009}, which are in fact based on counterfactual dependencies (do operator); our framework makes a clear distinction between (relativistic) causality and counterfactual dependencies in terms of time vs. hap.

So this is it for a 101 on special relativity. Now, let us move on to quantum physics.

\section{Quantum physics and possible worlds}

In quantum physics, a system can be in a state of superposition 
\cite{Dirac1947}, which is famously embodied in the Schrodinger's cat thought experiment \cite{Gribbin1985}, which attempts to bring superposition to a macroscopic level.

With current technologies, we are able to prepare and manipulate such states at the microscopic level, i.e., on the spin of one or several electrons \cite{Compton1921} or the polarization of photons \cite{Beth1935}. The size of the quantum systems we can build, which can be measured in qubits \cite{Schumacher1995} in quantum information and quantum computation, increases every year, even though this pace is slow and it also unknown whether there is a physical limit to what can be achieved.

A superposition, also called entanglement, is to be distinguished from a statistical mixture, in that it is not merely a statistical ensemble, but something more fundamental: statistical ensembles are called mixed states and correspond to the classical view of probabilities; they can notably be manipulated as density matrices \cite{Neumann1927} that have a rank greater than one. Pure states, although they are not mixed, can themselves be entangled. The evolution of pure states is mostly deterministic, and is governed by the Schrodinger equation \cite{Schrodinger1926} under classical physics, and the Klein-Gordon \cite{Gordon1926}  \cite{Klein1926} and Dirac \cite{Dirac1928} equations are two examples of generalization to the relativistic realm.

Mixed states, that is, classical probabilities, can be obtained through a mechanism that is irreversible\footnote{Although immersing the system undergoing a measurement process into a carefully chosen, larger one can provide a trick that makes the evolution reversible on the larger system. This is called a Stinespring dilation.}, and which is called measurement. Such systems can then be measured \cite{VonNeumann1958} by observing one of its properties, e.g., position, momentum, spin along some axis, up or down polarization, etc. The description of such measurements is, however, probabilistic and assigns a weight to each possible outcome, rather than always predicting the exact outcome of each single experiment. Performing the same experiment a large number of times allows us to confirm that the weights predicted by quantum theory, according to the Born rule \cite{Born1926}, match the actual frequency of the outcomes.

For example, if we prepare 1,000,000 qubits in the entangled state

$$\frac{\bra{0} + \bra{1}}{\sqrt(2)}$$

and perform a measurement in the computational basis, then we will typically \cite{Duerr2021} find around (not necessarily exactly) 500,000 times the result 0 and 500,000 times the result 1. Typicality means that it would be quite a big surprise\footnote{Formally (originally, Boltzmann, but also Dürr in the context of Bohmian mechanics), a ``big surprise'' means that an associated macroscopic variable would have a small volume in phase space; but over time, trajectories tend to go to bigger volumes, not smaller volumes. This is related to entropy and the second principle of thermodynamics, too.} to find, say, 1,000,000 times the result 0, even though theoretically possible, but in a typical world we would expect roughly a half-half distribution of the outcomes following the Born rule.

Hugh Everett \cite{Everett1957} suggested that, when we perform such a measurement, the universe forks, and each branch of the suchly obtained multiverse corresponds to one possible measurement outcome: a contingency\footnote{In modal logic, contingent means neither necessary nor impossible}. This is known as the many-worlds interpretation of quantum physics. Of course, the number of possible worlds is very high: alone with our previous examples, this would give us $2^{1000000}$ different worlds, a much higher number than the number of atoms in our visible universe.

The mainstream and classical view for picturing a many-world system is that of a tree structure, where each measurement forks the current world into several subtrees, one for each possible measurement outcome. This picture of a tree and forking universe for the measurement process is in fact strongly tied to the free choice assumption\cite{Colbeck2011}\cite{Wharton2020}\cite{Hance2022}\cite{Pawel2021}. This is because one formulation of free choice is that a freely chosen random variable located somewhere in spacetime, called an SV in literature \cite{Colbeck2011}, for spacetime variable, is independent\footnote{In the quantum foundations literature, independence in this context is meant as statistical independence, i.e., classical conditional probabilities factorize in a certain way given assumptions (d-separation in the causal dependency graph) implied by the causal structure, in which freely chosen variables are exogenous. As we will see though, in our view, what is truly the essence of free choice is counterfactual independence. Our framework makes it possible to rigorously distinguish between the two. We made an argument in \cite{Fourny2019} that statistical independence presupposes counterfactual independence. A short version of this argument is that if two random variables with full support are statistically independent from each other, then the joint random variable has full support on the Cartesian product of the underlying spaces, too; hence, the closest world in which a variable takes a different value is one in which all other variables have the same values.} of anything not within its future light cone.

To explain this picture, let us take for example one measurement of a system prepared in the diagonal state

$$\frac{\bra{0} + \bra{1}}{\sqrt(2)}$$

happening in a lab located somewhere in spacetime, and the physicist freely chooses either the original basis

$$(\bra{0}, \bra{1})$$

or the diagonal basis

$$\left(\bra{+}=\frac{\bra{0} + \bra{1}}{\sqrt(2)}, \bra{-}=\frac{\bra{0} - \bra{1}}{\sqrt(2)}\right)$$

Both the choice of measurement axis and the measurement outcome can be modeled with SVs $A$ and $O$, these two SVs corresponding to two timelike-separated events in spacetime.

Free choice of $A$ amounts to saying that there is nothing in the past or in the present (in the sense of everything spacelike separated) that contains any (classical) information from which we could predict the value of the chosen measurement axis, $A$, with certainty.

Furthermore, there are impossibility results (the Bell theorem \cite{Bell1964}, the Kochen-Specker theorem \cite{Kochen1975}, the free will theorem \cite{Conway2006}, the Colbeck-Renner theorem \cite{Colbeck2011} etc) that tell us that as a consequence of our axes being chosen freely\footnote{Note that the theorems involve more complex setups than a single measurement, such as EPR pairs of particles that are then sent far away from each other and measured independently. This is how what is called ``quantum contextuality'' is observed.}, the measurement outcomes are also free. In our case, there is also no information outside $O$'s future lightcone thanks to which we could predict the value of the measurement outcome $O$ with certainty. This is what makes quantum theory a probabilistic theory, with statistical predictions.

Another way of formulating it is that, under this mindset, after Alice chose the original axis, the world forks into a world in which $O=0$ and another in which $O=1$, these two worlds sharing the same past-and-present, where the past-and-present is made of all the values of all SVs in $O$ not in its future lightcone.

\begin{figure}
\includegraphics[width=\textwidth]{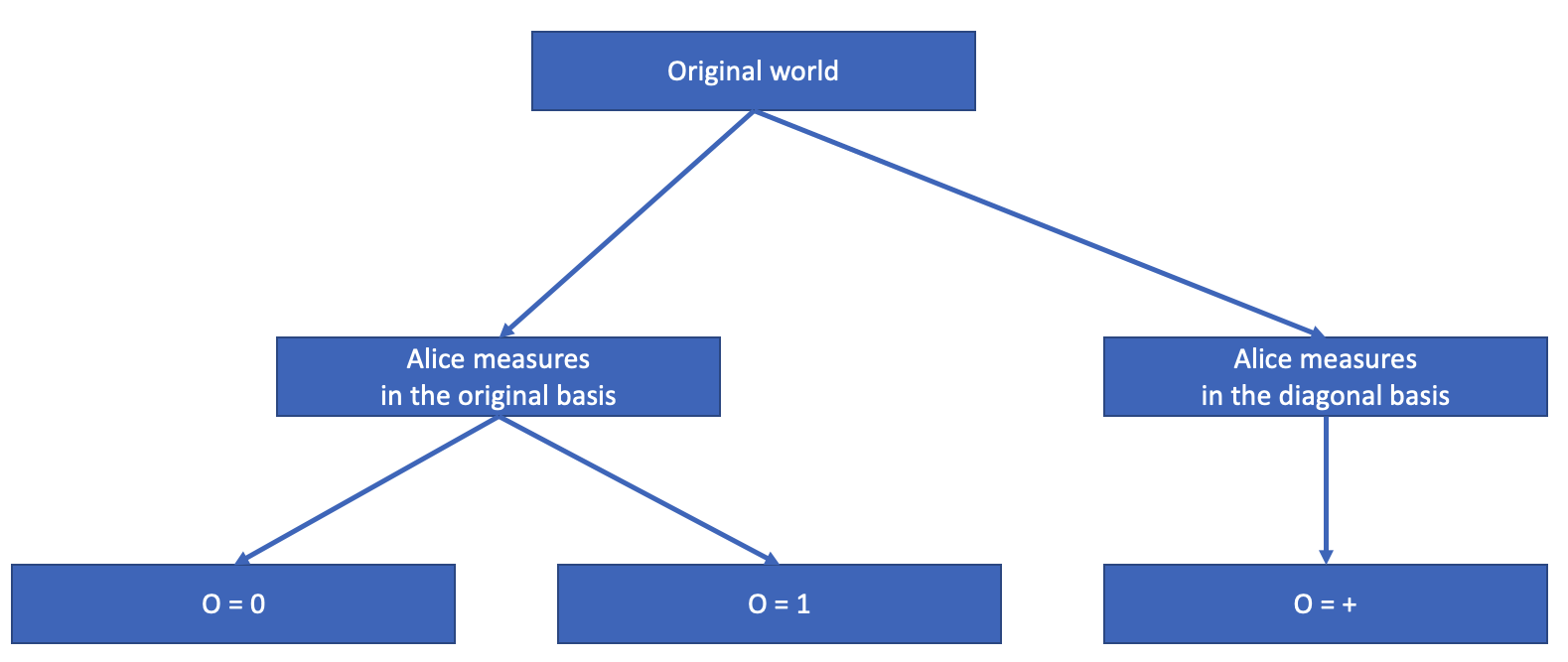}
\caption{A many-world structure, seen with the mainstream picture. Note that if Alice measures in the diagonal basis, the result is deterministic because the original state is already an eigenstate of the chosen observable, so there is no fork.}
\label{Fig1}
\end{figure}

This ``fork'' is what gives a tree structure to the common picture of the Everettian many-worlds semantics. We summarize the many world structure of our small thought experiment on Figure \ref{Fig1}.

However, there is another interpretation of quantum mechanics known as the De-Broglie-Bohm theory, or Bohmian mechanics \cite{DeBroglie1927}\cite{Bohm1952}. This theory models possible worlds as trajectories in configuration space ($\mathbb{R}^{3N}$ for N particles). Trajectories never cross each other, and each trajectory is fully defined by the initial position of the particles, i.e., a single point in configuration space. This theory is thus fully deterministic and makes the same predictions as quantum theory, although the problem of the unknown initial condition remains.

Under this picture, the possible worlds do not share the same past, because they correspond to disjoint trajectories from the beginning of the universe (the initial conditions). What is important to understand is that these trajectories are microscopic and cannot be observed directly with experiments: rather, quantum experiments and measurements involve a ``macroscopic interaction'', a bit similar to how thermodynamics looks at a gas through quantities such as temperature, energy and entropy but does not look at the position of single particles. Looking at a gas with a specific temperature and volume means looking at the ``ensemble'' of all the possible configurations that the particles can have under this temperature and volume. A possible microscopic configuration is called a micro-state, and the ensemble is called, in this case, the canonical ensemble.

Likewise, under the Bohmian view, making a measurement corresponds to a macroscopic variable. What happens with the many microscopic trajectories is that they fork in a process called ``decoherence'', many going down the $O=0$ path and many others going down the $O=1$ path. $O=0$ is then a macroscopic variable associated with the ensemble of trajectories that ``turned left'' while $O=1$ is associated with the ensemble of trajectories that ``turned right''.

\begin{figure}
\includegraphics[width=\textwidth]{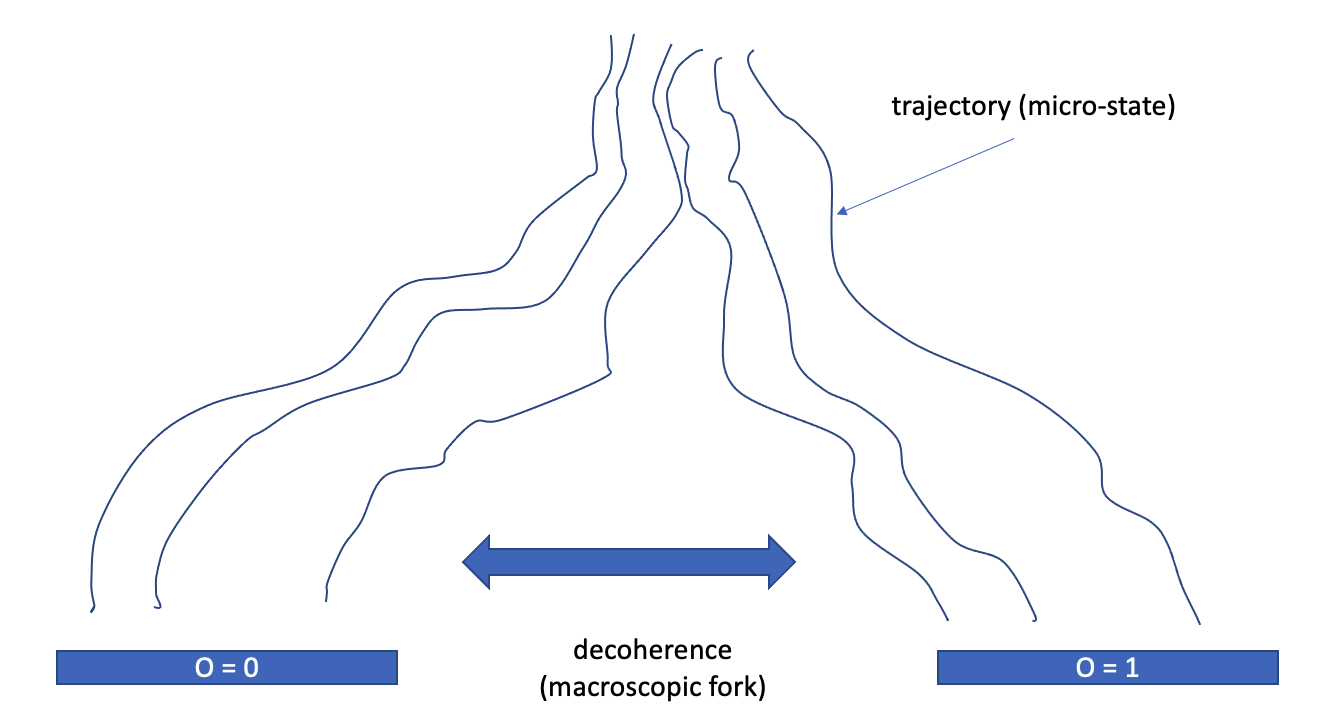}
\caption{Six Bohmian trajectories, that fork into two subtrees of three each. Time is represented vertically top-down, and configuration space horizontally (in just one dimension to simplify). In reality, there are also trajectories ``inbetween'', however their probability density is so low that they can be considered negligible (atypical) and, for all practical purposes, the trajectories fork into two subsets.}
\label{Fig2}
\end{figure}

As decoherence patterns follow each other, the Bohmian trajectories keep forking from each other in smaller and smaller ensembles, which means that the tree structure is in fact ``macroscopically visible'' in the configuration space evolving through time, while all the information about future measurement outcomes is still contained at the microscopic level in each trajectory.

\begin{figure}
\includegraphics[width=\textwidth]{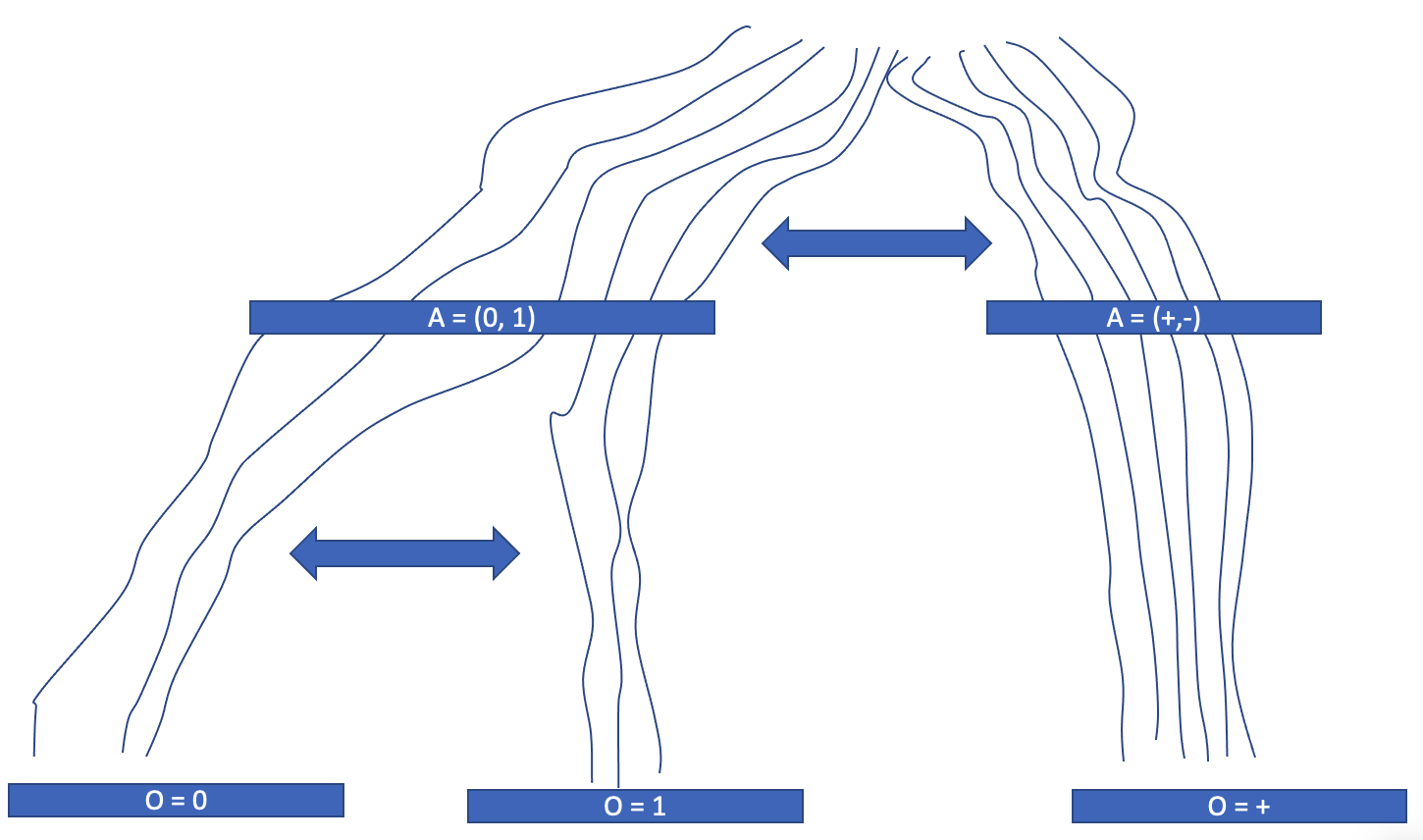}
\caption{Twelve Bohmian trajectories corresponding to the macroscopic view shown on Figure \ref{Fig1}. Time is represented vertically top-down, and configuration space horizontally (in just one dimension to simplify). In reality, there are also trajectories ``inbetween'', however their probability density is so low that they can be considered negligible (atypical) and, for all practical purposes, on the macroscopic level, the trajectories fork into these three subsets.}
\label{Fig3}
\end{figure}

Figure \ref{Fig3} shows trajectories that could correspond to the many-world tree structure shown on Figure \ref{Fig1}. The reader should note that we did something not typically found in Bohmian literature: we also considered the human choice of the measurement axis to form a decoherence pattern in Bohmian trajectories, while most of the Bohmian literature keeps this choices outside of the trajectories. It is our belief that choices of measurement axes and measurement outcomes are physically the exact same phenomenon, just like the Moon orbiting the Earth and a ball bouncing on the Earth surface are due to the same gravitational effects.

Whether alternate possible worlds are real or not, according to either the many-world interpretation or Bohmian mechanics, is the subject of intense debate. But irregardless of this, even if we considered that they are not real, they are nevertheless part of these theories as alternate contingencies, as ``what could have been''. This is true even with interpretations involving a wave-function collapse, like the Copenhagen interpretation: one measures the spin up, but \emph{it could have been that we measured it down instead}. The tree structure of worlds also exists in the Cophenhagen interpretation as mere theoretical possibilities.

The degree of accuracy with which quantum theory makes predictions beyond the reach of classical theories is thus a strong indicator that contingency plays an important role in the laws of our universe.

This in turn directly suggests that an event is not only identified, by an observer, with its location in space and in time, but is also identified by the possible world in which it occurs. We call the latter a \emph{happenstance} coordinate, or \emph{hap} coordinate to keep it short. In a theory akin to Bohmian mechanics, this third set of coordinates is a point in configuration space that is the initial condition corresponding to the actual world.

In other words, while we view our own (actual) world as four-dimensional space-time, we suggest that the entire multiverse should be formally modelled as a space-time-hap manifold, where \emph{hap} is a different kind of dimension, and is complementary to space and to time.

Note that this model does not take sides on whether other worlds than the actual world are part of reality, which is an entirely orthogonal, philosophical discussion. This is similar to modelling probabilities with an $\Omega$ space with one elementary event being the actual world and all other elementary events in the $\Omega$ space being hypothetical.

Given an observer, each event thus has a location in time, a location in space, and a location in hap. We now formally propose a coordinate system to expand spacetime with hap locations.

\section{Assumptions on the underlying De-Broglie-Bohm-like theory}

In this note, we use the expression ``De-Broglie-Bohm-like'' for several reasons. First, the vanilla description of the De-Broglie-Bohm theory, which uses the Schrodinger equation, is not covariant and thus not considered compatible with special relativity. An extension to special relativity was provided only for one particle, and has not been contributed yet for more particles. It has also been suggested that there could be a so-called \emph{preferred foliation}, i.e., a preferred reference frame in which the trajectories are computed (see \cite{Duerr1999} for an argument, and \cite{Galvan2015} for a counterargument). The present note is orthogonal to these consideration and aims to be forward compatible with any similar future theory, under a few simple and abstract assumptions including that:

\begin{itemize}
\item there exists a configuration space, which we expect to be of dimension $3N$ in this note, but this can be replaced with any other appropriate number;
\item a possible set of trajectories consists in (not necessarily straight) lines in the vector space obtained by extending configuration space with one dimension of time (``configuration spacetime'', which we will shortly be calling haptime for reasons we will explain);
\item the trajectories in a possible set of trajectories never cross -- in other words, the trajectories form a foliation of ``configuration spacetime'' (we will be shortly calling it a haptime foliation) with leaves of dimension 1.
\end{itemize}

Under these conditions, a possible set of trajectories can be characterized with a family of (bijective) guiding functions $(G_{t=t_1 \rightarrow t=t_2})_{t_1, t_2}$ that, given two points in time, move, a particle along a trajectory from position $c_1$ at $t_1$ to its new position $c_2$ at time $t_2$ following the guiding equation of the underlying theory:

$$c_2= G_{t=t_1 \rightarrow t=t_2}(c_1)$$

Note that the inverse of $G_{t=t_1 \rightarrow t=t_2}$ is $G_{t=t_2 \rightarrow t=t_1}$.

Note that a change of observer translates naturally to a change of coordinates $(t, c)$ to $(u, d)$ in configuration spacetime:

$$\forall i, (u, d_i)=B_\textbf{v}(t, c_i)$$ 

where $B_\textbf{v}$ is the corresponding Lorentz boost, and the same set of trajectories represented by $G$ translates as well to a different guiding equation according to this other observer. We can thus naturally extend the notation to this different coordinate system, with the same set of trajectories\footnote{One could (abusively) say that $G$ is tensorial, or Lorentz invariant in the sense that a set of trajectories can be equivalently expressed in any coordinate system. However, it is important to understand that this does not imply that all observers agree on which set of trajectories they consider to be the ``truth'' even up to a Lorentz transformation.}:

$$d_2= G_{u=u_1 \rightarrow u=u_2}(d_1)$$

Seeing that $t=t_1$ is generally a spacetime hyperplane equation (a time hyperplane in the coordinate system $(t, x)$), we can generalize moving along the configuration trajectory to guide any particle to any hyperplane, for example with the coordinate system $(u, y)$ for another observer, like so:

$$d_1=G_{t=t_1 \rightarrow u=u_2}(c_2)$$

where $d_1$ is expressed with the coordinate system of the other observer. Noting that this is only a convenient shortcut notation that embeds the Lorentz Boost for each particle.

If there is not a preferred foliation, but several sets of trajectories may coexist for different observers, we use superscripts ($G^A$, $G^B$...) for different sets of trajectories.

\section{A coordinate system for spacetimehap}

We formally describe the space-time-hap as a vector space in the specific case of (i) flatness (masses and gravity are considered negligible so that the rules of special relativity apply) and (ii) a constant number N of particles (no creation or annihilation). I believe it is possible to extend the construction to a variable number of particles using Fock spaces, but leave it out of scope for this note. Likewise, I believe the theory should naturally extend to curved spacetimehap but assuming flatness simplifies the discussion.

\begin{definition} A flat space-time-hap manifold is a vector space $\mathbb{R}^{1+3+3N}$ where:
\begin{itemize}
\item three dimensions represent space
\item one dimension represent time
\item $3N$ dimensions represent hap
\end{itemize}
and in which each observer (inertial reference frame) uses their own basis.
\end{definition}

The $3N$ dimensions can be interpreted as a configuration space, in that the hap dimensions correspond to the positions of the $N$ particles in a De-Broglie-Bohm-like theory as seen by the specific observer at time $t=0$. Indeed, in the De-Broglie-Bohm theory, these initial conditions uniquely determine the trajectory jointly followed by the particles in configuration space and are a natural identifier for the actual world.

Note that a direct consequence of the above assumptions is that, for a specific observer located at a position with time $t_O$, each point in configuration space identifies exactly one trajectory, and conversely each trajectory is uniquely identified with a point in configuration space, which is its intersection with the $t=t_O$ hyperplane for the said observer.

Thus, the configuration space of the (open) underlying De-Broglie-Bohm-like theory naturally acts as the hap in our spacetimehap manifold.

Now that we have our flat spacetimehap manifold defined as a vector space, given an observer or equivalently a given inertial frame, we can derive a coordinate system in a straightforward way.

\begin{definition} A spacetimehap event $E$ is a point $(t_E, x_E, c_E)$ in the vector space $\mathbb{R}^{1+3+3N}$ where:
\begin{itemize}
\item $t_E\in\mathbb{R}$ is the point in time at which the observer sees (or would see) the event happen
\item $x_E\in\mathbb{R}^3$ is the position in space at which the observer sees (or would see) the event happen
\item $c_E\in\mathbb{R}^{3N}$ is the position in hap, and embodies the initial conditions under which the event happens
\end{itemize}
\end{definition}

The spacetimehap event $E$ with coordinates $(t_E, x_E, c_E)$ can be interpreted like so: ``The initial positions of all $N$ particles at $t=t_E$ are $c_E$, the time is $t_E$ and the position is $x_E$.'' Coordinates in space and time identify an event within one possible world, while the hap coordinates identify the world in which the event takes place within the set of all possible worlds.

We now generalize the concept of timelike separation and spacelike separation to spacetimehap. In spacetimehap, any two events can be either timelike-separated, or spacelike-separated, or haplike-separated.

The main visual to have is that each possible world $\omega$, identified with some hap-coordinate $c_{\omega}$, is a dimension-4 (affine) subspace of spacetimehap that is the section of spacetimehap with equation of $c=c_\omega$. If one then forgets about the (unique) hap coordinate of this world, it is simply a regular Minkowski spacetime manifold within which events (that all have $c=c_\omega$ as their hap coordinates) can be timelike-separated or spacelike-separated.

Under a De-Broglie-Bohm-like theory, a combination of equations, for example the Schrodinger equation and the guiding equation, lead to a foliation of haptime of which the leaves are the trajectories.

Under the assumption that no particle can travel faster than light, there are cases in which two pairs of events can in no circumstance ever be in the same world, and this, no matter what the underlying equations are. This happens in particular if the spacetime position of one of the particles in the first event (given by the projection on the time coordinate, and the corresponding three coordinates of hap, which are akin to space for this one particle) is spacelike-separated from its spacetime position in the second event.

\begin{definition}[Haplike separation] Given two spacetimehap events $E_1=(t_1, x_1, c_1)$ and $E_2(t_2, x_2, c_2)$, they are said to be haplike-separated, if

$$\exists p \in [1..N] \text{ s.t. } \|(t_1-t_2, c_1^P-c_2^P)\|^2 < 0$$
\end{definition}

If two events are haplike separated, then they are always counterfactuals with respect to each other, i.e., these events cannot happen in the same possible world.

Note that the above definition is Lorentz-invariant and is thus independent of the inertial reference frame.

When two events are not haplike separated, there is a possibility, depending on the equation, that they might happen in the same world. This depends on the specific haptime foliation given by the theory. For theories without a preferred foliation, this might be also depend on the observer.

Events that are not haplike separated can be spacelike-separated or timelike-separated\footnote{We consider in this note spacelike separation strictly and timelike separation not strictly, i,.e., null-separated is considered timelike-separated, because we are focused on causal semantics. This convention might differ from other papers that define timelike-separation as chronological separation and not causal separation.}. The definition of spacelike-separation and timelike-separation of two spacetimehap events is simply the same as in Minkowski spacetime, considering the projection of the spacetimehap events unto their spacetime coordinates and ignoring their hap part.

\begin{definition}[Timelike separation] Given two spacetimehap events $E_1=(t_1, x_1, c_1)$ and $E_2(t_2, x_2, c_2)$, $E_1$ is said to causally precede $E_2$, and the events are said to be (future-directed) timelike-separated, if:
\begin{itemize}
\item $E_1$ and $E_2$ are not haplike-separated
\item $(t_1, x_1)$ causally precedes $(t_2, x_2)$ under Minkowski spacetime semantics (including the case when the spacetime distance is 0).
\end{itemize}
\end{definition}

\begin{definition}[Spacelike separation] Given two spacetimehap events $E_1=(t_1, x_1, c_1)$ and $E_2(t_2, x_2, c_2)$, $E_1$ and $E_2$, and the events are said to be spacelike-separated, if:
\begin{itemize}
\item $E_1$ and $E_2$ are not haplike-separated
\item $(t_1, x_1)$ and $(t_2, x_2)$ are spacelike-separated under Minkowski spacetime semantics (excluding the case when the spacetime distance is 0).
\end{itemize}
\end{definition}

Note that timelike-separation, spacelike-separation, and haplike-separation are mutually exclusive and complementary. In particular, two haplike-separated events are never spacelike-separated nor timelike-separated even though it would be tempting to say they are by only looking at their spacetime coordinates.

Let us now come back to our observer, who is associated with a given inertial frame of reference with respect to space and time. Since we extended the spacetime framework with hap, our observer is not only located at a specific position in spacetime and moving at a certain (relative to other inertial frames) constant speed; our observer $O$ is also located in a specific possible world, one of the leaves of the haptime foliation. As a consequence, from their perspective, events are to be classified in two categories: \emph{actual events} who are in the same leaf, and \emph{counterfactual events} that are in a different leaf. From a terminology perspective, actual events actually \emph{happen}, or \emph{have actually happened}, or \emph{will actually happen} (mind the indicative mode), while counterfactual events \emph{could happen} or \emph{could have happened}. Do mind the subjunctive mode, as this terminology is more precise than talking about counterfactuals in (casual) terms of ``changing'' something, which is the source of much confusion in the literature: one does not ``change'' a measurement axis; one considers what \emph{would have happened if the measurement axis had been chosen differently}.

With this framework, it is this possible to formally discuss mixed states from the perspective of an observer. For example, if we consider a measurement against the computational axis leading to a statistical mixture of the ensemble ($\ket{0}$, $\ket{1}$), then Alice may observe the event of measuring $\ket{0}$ as an actual measurement outcome (``I have measured 0.'') located at a timelike-separated position (her past), and for her $\ket{1}$ is a counterfactual (``I could have measured 1.'') located at a haplike-separated position; and Bob (haplike-separated from Alice) will observe the event of measuring $\ket{1}$ as an actual measurement outcome (``I have measured 1.'') located at a timelike-separated position (his past), and for him $\ket{0}$ is a counterfactual (``I could have measured 0.'') located at a haplike-separated position.

Thus, spacetimehap provides a framework for the formal interpretation of mixed states with a many-world flavour of quantum mechanics.

\section{Change of coordinates}

We now describe how the coordinates change with a change of the frame of reference.

Let us assume that we have an observer Alice, and that she sees another observer Bob moving at a constant speed $\textbf{v}$ from her.

Note that some theories might allow a situation in which Alice sees Bob in her world, but Bob sees Alice in a different world if the haptime foliation is observer-dependent. For now, however, we will assume a theory with a preferred, observer-independent haptime foliation, and assume in particular that Alice and Bob are located in the same world and both agree with this.

Let us assume Alice sees an event $E$ at coordinates $(t_E, x_E, c_E)$. Concretely, this event represents what, from her standpoint, she \emph{would have seen} at the spacetime coordinates $(t_E, x_E)$ if the initial conditions had been $c_E$ at time $t_E$.

Bob sees this event in a different coordinate system $(u_E, y_E, d_E)$ where:

\begin{itemize}
\item $(u_E, y_E)$ are obtained from $(t_E, x_E)$ via a Lorentz transformation corresponding to speed $\textbf{v}$.
\item $d_E$ corresponds to the positions of the $N$ particles at $u_E$; these can be uniquely computed from $c_E$ and $\textbf{v}$ by (i) considering the haptime foliation leaf to which $(t_E, c_E)$ belongs and then (ii) intersecting this leaf with the haptime hyperplane $(u_B)$ where the hyperplane is considered from Bob's perspective.
\end{itemize}

Let us dive more into the change of coordinates for $c_E=(c_{E,p})_{1\leq p \leq N}$.

There are two possible reasoning paths to reproduce the geometric description of the intersection of the haptime foliation leaf with the hyperplane.

In one reasoning, we consider the position of all particles seen by Alice at time $t_E$ according to $c_E$, which is the family of spacetime coordinates $(t_E, c_{E,p})_p$. We then use a Lorentz transformation on these spacetime events to get their coordinates in Bob's reference frame, i.e., for each $p$:

$$(u'_{E,p}, d'_{E,p}) = B_\textbf{v}(t_E, c_{E,p})$$

The difficulty here, however, is that we get initial conditions that have a different time for each particle according to Bob's reference frame. So we need to move each particle following the guiding equation to bring it to a spacetime position that Bob sees at time $u_E$. Using the notation previously introduced to express the guiding equation across several coordinate systems, this gives us the hap coordinate as seen by Bob:

$$d_E=G_{t=t_E \rightarrow u=u_E}(c_E)$$

This contains the Lorentz boost, which can be interpreted in one of to ways depending on who ``sees'' the guiding equation:

\begin{enumerate}
\item The Lorentz Boost is first applied to each particle as seen on the $t=t_E$ hyperplane, to positions in Bob's coordinate system that have different times. Then, the guiding equation (seen by Bob) is used to bring them all the way to $u=u_E$.
\item The guiding equation (seen by Alice) is first used to bring each particle $p$ to a carefully picked time $t_p$ and position $x_p$ in such a way that the Lorentz boost applied to $(t_p, x_p)$ gives $u=u_E$.
\end{enumerate}

Note that the two ways are equivalent if Alice and Bob see the same set of trajectories (up to their Lorentz boost), i.e., if there is a preferred haptime foliation.

\section{Can the haptime foliation be observer dependent?}

The reason why this second line of reasoning is relevant becomes clear in the case that we do not assume a preferred haptime foliation: different observers might disagree on whether two spacetimehap events happen in the same world or not. This means that, from Alice's perspective, two events might happen in the same world, while for Bob, they may be counterfactual to each other. This relaxation is akin to the early insights brought by relativity theory: that whether two spacetime events happen at the same time or not may depend on the observer.

It should be said, however, that our definition of haplike separation implies that two haplike-separated events will always be seen as mutually counterfactual for any observer: nobody will see them happen in the same possible world. This is also akin to the definition of the timelike separation of two events, which will never be seen as concomitant\footnote{happening at the same time} by different inertial frame observers. 

The framework introduced in this note allows for future avenues of research on the consequences of making this relaxation. In particular, without preferred haptime foliation, the two lines of reasoning in the previous section may lead to different changes of coordinates, meaning that this change is \emph{asymmetric}. If Alice sees the set of trajectories $G^A$ and Bob sees the set of trajectories $G^B$, then this leads to two different changes of coordinates:

$$d_E=G^A_{t=t_E \rightarrow u=u_E}(c_E)$$

and

$$d_E=G^B_{t=t_E \rightarrow u=u_E}(c_E)$$

\section{Free choice is observer-dependent}

Free choice is commonly described in game theory as a unilateral deviation. In our framework, this corresponds to a deviation of position of just one particle with all particle positions remaining the same.

Note that such a deviation is not a \emph{change} of position in the sense that the particle is moving. Rather, it is a subjunctive conditional that can be formulated as ``if the particle had been here instead at that time, and all other particles had been at the same positions as they are, then...''.

We take the view here that there is an observer-dependent haptime foliation, i.e., all observers see the same trajectories (G).

Without loss of generality, we assume that it is the first particle that is moving unilaterally for Alice, at $t=0$.

Now we consider whether such a deviation being unilateral depends on the position of the observer in spacetime. 

Concretely, this means that given two possible initial conditions $c_E$ and $c'_E$ seen by Alice and such that

$$c_{E,1}\neq c'_{E,1}$$

$$\forall i \text{ s.t. } 2 \leq i \leq N, c_{E,i}=c'_{E,i}$$

we are asking whether it is also true for Bob that 

$$d_{E,1}\neq d'_{E,1}$$

$$\forall i \text{ s.t. } 2 \leq i \leq N, d_{E,i}=d'_{B,i}$$

where $d_E$ and $d'_E$ are obtained from $c_E$ and $c'_E$ according to the change of coordinates described previously:

$$d_E=G_{t=t_E \rightarrow u=u_E}(c_E)$$

$$d'_E=G_{t=t_E \rightarrow u=u_E}(c'_E)$$

We see that these conditions bring additional, non-trivial constraints to $G_{t=t_E \rightarrow u=u_E}$, and that it can at most hold for the highly specific scenarios in which the above equalities are obtained.

It is helpful, for a more detailed analysis, to take the case of two particles moving in one dimension. Then there are two hap dimensions, one space dimension and one time dimension. The haptime foliation corresponding to G can then be visualized in three dimensions, with the coordinate system ($c_1$, $c_2$ and $t$) from Alice's perspective.

If we keep Alice's perspective and start with hap coordinates $c_E$ at $t_E$ and then follow the trajectory to another time $t_F$:

$$(c_{F,1}, c_{F,2})=G_{t=t_E \rightarrow t=t_F}(c_{E,1}, c_{E,2})$$

And now consider a perturbation with only the first particle's initial position at $t_E$, and look at how it propagates to $t_F$:

$$(c_{F,1}+\delta c_{F, 1}, c_{F,2} + \delta c_{F, 2})=G_{t=t_E \rightarrow t=t_F}(c_{E,1} + \delta c_{E, 1}, c_{E,2})$$

If we now look at the intersection of the set of trajectories with the $u=u_E$ hyperplane in Bob's coordinate system, we get:

$$(d_{E,1}, d_{E,2})=G_{t=t_E \rightarrow u=u_E}(c_{E,1}, c_{E,2})$$

Next, we can adjust the choice of Alice's time to $t=t_H$  such that $(t_H, c_{H,2})$ is seen to be at $u_E$ by Bob. Thus, the coordinates are related via the Lorentz boost:

$$(u_E, d_{E,2})=B_\textbf{v}(t_H, c_{H,2})$$

And the corresponding perturbation propagating to $t=t_H$ is:

$$(c_{H,1}+\delta c_{H, 1}, c_{H,2} + \delta c_{H, 2})=G_{t=t_E \rightarrow t=t_H+ \delta t_H}(c_{E,1} + \delta c_{E, 1}, c_{E,2})$$

$$(u_E, d_{E,2}+\delta d_{E, 2})=B_\textbf{v}(t_H+\delta t_H, c_{H,2}+\delta c_{H, 2})$$

The free choice constraint translates, for Bob, to:

$$(u_E, d_{E,2})=B_\textbf{v}(t_H+\delta t_H, c_{H,2}+\delta c_{H, 2})$$

meaning

$$B_v(t_H, c_{H,2})=B_\textbf{v}(t_H+\delta t_H, c_{H,2}+\delta c_{H, 2})$$

Knowing that

$$B_\textbf{v}(t, c)=(\gamma(t + v c), \gamma(c + v t))$$

We get

$$B_\textbf{v}(t_H+\delta t_H, c_{H,2}+\delta c_{H, 2})=(\gamma(t_H+\delta t_H + v (c_{H,2}+\delta c_{H, 2})), \gamma(c_{H,2}+\delta c_{H, 2} + v (t_H+\delta t_H)))$$

and

$$B_\textbf{v}(t_H, c_{H,2})=(\gamma(t_H + v c_{H,2}), \gamma(c_{H,2} + v t_H))$$

which we can subtract from each other:

$$0 =(\gamma(\delta t_H + v \delta c_{H, 2}), \gamma(\delta c_{H, 2} + v \delta t_H))$$
 
$$0 =(\delta t_H + v \delta c_{H, 2}, \delta c_{H, 2} + v \delta t_H)$$
 
That is

$$\delta t_H = -v \delta c_{H,2}$$
$$\delta t_H = -\frac{1}{v} \delta c_{H,2}$$

Which is only possible, if the perturbations are non zero, with $v=1$, i.e., Alice and Bob's frames of references differ by the speed of light, which cannot be.

Thus,

$$\delta t_H = 0$$
$$\delta c_{H,2} = 0$$

Which we put back into the guiding equation:

$$(c_{H,1}+\delta c_{H, 1}, c_{H,2})=G_{t=t_E \rightarrow t=t_H}(c_{E,1} + \delta c_{E, 1}, c_{E,2})$$

This constraint on the guiding equation can be visually seen as the absence of torsion in the set of trajectories, which is a non-trivial constraint. In fact, it comes down to the two particles being independent from each other, i.e., the global wave function is the simple product of the wave functions of the two particles.

Conversely, any entanglement of the two particles, which is the general case, implies that free choice is observer-dependent: while Alice would see a perturbation only on the first particle (a ``free choice'' of its position), Bob would see a perturbation on both, i.e., would not see the first particle ``moving freely." In counterfactual terms, the position of the first particle is not counterfactually independent from the position of the second one.

\section{Time, Space, Hap and the three kind of dependencies}

In this section, we relate each one of the kinds of dimensions in spacetimehap to kinds of dependencies in order to insist that this framework facilitates the distinction between them, as a dependency triangle. It is surprising to observe that, while most are aware that causal dependencies are different from correlations, a conflation is commonly made
\begin{itemize}
\item on the one hand between causal dependencies and counterfactual dependencies, by the very definition of (not relativistic) causal dependency!)
\item and on the other hand between counterfactual dependencies and correlations (which leads many to believe that the laws of conditional probabilities are different in the quantum world -- but we think that these are not, in fact, conditional probabilities, these are probabilities of subjunctive conditionals, which obey different laws, and a different notation should be used.
\end{itemize}

\subsection{Hap and counterfactual dependencies}

The hap dimension corresponds to counterfactuals, i.e., hap coordinates refer to the coordinates of the possible world in which a spacetimehap event takes place. Under a De-Broglie-Bohm-like theory, for a given observer, the wave function is a complex scalar field on hap and, through its squared norm, implies a probability measure on the hap coordinates at any particular time. Moreover, for a specific observer, if there is quantum equilibrium, the probability assigned to any given world is independent from the moment of time chosen, because the equation guiding the deterministic evolution of the wave function (Schrodinger or otherwise) preserves hap volumes.

There are several open problems as of 2022: there are theories that do not guarantee quantum equilibrium (this is a feature commonly encountered in theories that aim at being Lorentz-invariant), or that only do so for a given observer (i.e., under a preferred spacetime foliation).

Another interesting question is whether there exists any theory that is both Lorentz-invariant and has quantum equilibrium, i.e., the haptime foliation does not depend on the observer.

In the case where we consider a specific haptime foliation and under quantum equilibrium (either for a specific observer, or observer-independent in a Lorentz-invariant theory), i.e., there is a probability measure on possible worlds or, equivalently, on the hap coordinates at some specific time. One can then consider endowing the hap coordinates with a metric tensor (if it exists, which we leave as an open problem) in such a way that it is compatible with the probability measure, that is, such that the canonical Borel measure of the metric tensor matches the probability measure.

An immediate implication is that the volume of hap at any specific time must be finite, because the probability measure is normalized to one\footnote{Although one could also consider letting the hap volume diverge, akin to what is commonly accepted (and debated) in path-integrals.}.

Another immediate implication is that one can use the metric tensor to measure the arc length between two different hap coordinates at a specific time. This means that we can define a distance $d$ between any two worlds. An open question and challenge is whether, under quantum equilibrium, it is possible to synchronize the metric tensors at every moment in time in such a way that the arc length is preserved across time, similar to the preservation of volume.

In turn, this distance can be used to define the function $f$ underlying counterfactual dependencies in decision theory on the set of all possible worlds $\Omega$, which is the leaf of the haptime foliation obtained from the guiding equation: given a world $\omega\in \Omega$, identified with hap coordinates at some moment in time, some random variable A with full support and a value $a$ in its target set, we can define $f_{A=a}(\omega)$ as the set of the worlds with a given distance from $\omega$ and in which $A=a$, the given distance being the mimimum distance so that the set is non-empty.

\subsection{Time and causal dependencies}

Not surprisingly as this is common in the literature, timelike-separation between SVs can be used to define relativistic causality and to abstract it away with a causal dependency graph where each node is an SV and each directed edge corresponds to future-directed timelike separation between two SVs. We furthermore consider the transitive reduction of this graph.

There are many models and approaches viewing causal dependencies with the angle of a causal dependency graph in the literature: causal models \cite{Pearl2009}, quantum causal models, from a decision-theoretical perspective.

\subsection{Space and statistical dependencies}

A common characteristic of quantum effects is that they break Bell inequalities in the EPR experiment; formally, the set of possible conditional probabilities $p_{XY|AB}(x, y, a, b)$ where x, y (outcomes), a and b (axes)\footnote{This is the Zurich convention, the Geneva convention is different} take the values 0 or 1 can be seen as a dimension-16 hypercube, and Bell inequalities as a simplex subregion of this hypercube that constrains local, realist theories under free choice.

It is important to pause for a moment and make explicit how these conditional probabilities are obtained. In a laboratory, the same experiment is repeated, many times, with various choices of axes, and data is accumulated and recorded. Then, correlations are computed from this data. In particular, all these records are in the same world, and appear jointly on the same piece of paper: they are spacelike-separated.

This is how correlations (spacelike) differ from counterfactual dependencies (haplike): when we ask what would have happened if we had done the same experiment with a different choice of axes, we are not actually recording this other hypothetical experiment on the same piece of paper as the actual experiment, but on some other piece of paper in a parallel world (haplike-separated); correlations are thus inadequate to describe this setup and counterfactual dependencies should be used instead. We thus believe that hapspacetime with its three kinds of dimensions can be useful for gaining new insights and more precision in discussions around contextuality and in particular the Kochen-Specker theorem  \cite{Kochen1975} and Mermin's squares \cite{Mermin1990}. With this new perspective, locality can thus be assumed in weaker terms of timelike-separation only (as opposed to counterfactual dependencies across hap, which do not contract locality under the theory of relativity), realism in weaker terms of values preexisting their measurements in the actual world (as opposed to counterfactuals on different, haplike-separated measurements), and free choice in weaker terms of full support (all possible values are found across hap), with these three weaker forms of locality, realism and free choice jointly compatible with what we observe (they allow breaking Bell inequalities) \cite{Baczyk2022}.

\section{Generalized ergodicity in spacetimehap}

Our framework also allows for an observation related to thermodynamics and ergodicity.

In classical thermodynamics, one can consider the evolution of the speed (or position) of a single particle over time, or one can consider all speeds (or positions) of all particles in a gas at the same time and ergodicity tells us that the distributions are identical.

This observation can be generalized to spacetimehap in the following way

\begin{itemize}
\item (time, causal) the position of a single particle, in a specific world, over the entire time
\item (space, statistical) the position of all particles, in a specific world, at a specific time
\item (hap, counterfactual) all possible (in the sense of in all possible worlds) positions of a single particle at a specific time, i.e., the actual position, as well as all the other hypothetical positions at which this particle \emph{could have been}.
\end{itemize}

A generalized ergodicity principle would thus state that the three above distributions are equivalent.

\section{Declarations}

\subsection{Funding} The paper was written under the author's affiliation to ETH Zurich. There was no other funding.

\subsection{Conflicts of interest/Competing interests} There are no conflicts of interest or competing interests to declare.

\subsection{Availability of data and material (data transparency)}  Not applicable as this is a theoretical contribution.

\subsection{Code availability}  Not applicable as this is a theoretical contribution.

\subsection{Data availability statement} There is no underlying data as this is a theoretical contribution.

\subsection{Authors' contributions}  Not applicable as there is only one author.


\bibliographystyle{spbasic}      

\end{document}